\documentclass{PoS}

\title{Experimental Overview of Light Mesons}

\ShortTitle{Experimental Overview of Light Mesons}

\author{\speaker{Sean Dobbs}%
         \thanks{Previous affiliation: Northwestern University.}\\
        Florida State University\\
        E-mail: \email{sdobbs@fsu.edu}}

\abstract{The light quark $uds$ mesons have been a foundation of our understanding of the strong interaction for decades. New experiments with modern detectors and large data sets are furthering our understanding of the spectrum and dynamics of these states.  In this paper, I review several recent results on the spectroscopy of light mesons and discuss the future of this field.}

\FullConference{XVII International Conference on Hadron Spectroscopy and Structure\\
                 25-29 September, 2017\\
                 University of Salamanca, Salamanca, Spain}

\begin{document}

\section{Introduction}

The mesons are bound states primarily consisting of a quark and anti-quark, and are the simplest strongly interacting systems.  We have a theory of the strong interaction: Quantum Chromodynamics (QCD). However, the development of a detailed understanding of the properties of QCD in the strongly coupled regime has so far been incomplete, particularly with regards to its physical states.  The study of the properties and decays of these states, particularly of the mesons, is expected to shed light on some of the open questions in QCD.  These questions include which color-singlet quark states are allowed by nature, and what is the detailed nature of the interaction responsible for quark confinement.  Recent advances in experiment and theoretical understanding have allowed to continue to move forward towards answers to these questions.

The light mesons composed of $uds$ quarks are particularly interesting for two reasons. Due to their light mass, they probe relativistic effects and the confinement regime of the quark-antiquark interaction more strongly than those containing heavier quarks. We also have the ability to experimentally generate them in copious amounts.  In this paper, I will only discuss light-quark mesons that are ``non-exotic'', i.e., appear to be primarily composed of a quark-antiquark pair.  Heavier quarks and more exotic states are covered by other papers from this conference.  Also, even this more restricted topic is incredibly rich, so I will only focus on discussing a few of the most prominent latest experimental results in the spectroscopy of these light mesons.

Mesons are most simply understood in terms of the constituent quark model, as a valence quark and anti-quark pair.  They are classified by their quantum numbers, total spin $J$, parity $P$, and charge conjugation $C$.  The light mesons are found to come in ``octets'' of mesons with the same $J^{PC}$.  The spectrum of these states can be calculated using phenomenological models or numerical calculations of QCD on a lattice (LQCD)~\cite{lattice}, and are broadly found to be in good agreement with states that have been observed so far.  In the latest edition of the PDG, over 80 mesons have been identified~\cite{pdg}.  In the most general sense, we can say that most lowest state octets are well-established (with some notable exceptions, such as the the $0^{++}$ scalar mesons), while the excited states in the range $M>2$~GeV are less understood.  It is fair to say that we will not truly understand the spectrum of light mesons unless we can understand the radial excitations of the lowest mass mesons.  Also, since several types of exotic states are expected to lie in this higher mass range, the firm establishment of the spectrum of light mesons has become important not just for the intrinsic understanding of these states, but as a prerequisite for the study of these exotic states.

\begin{figure}[!tb]
\begin{center}
\includegraphics[width=2.1in]{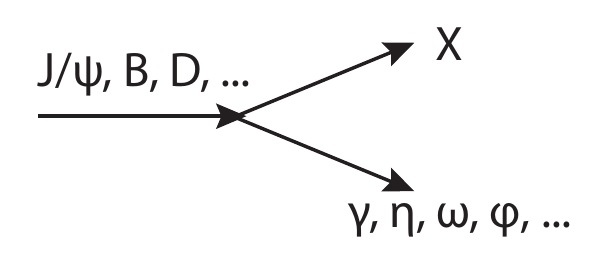}
\hspace{-10pt}
\includegraphics[width=2.4in]{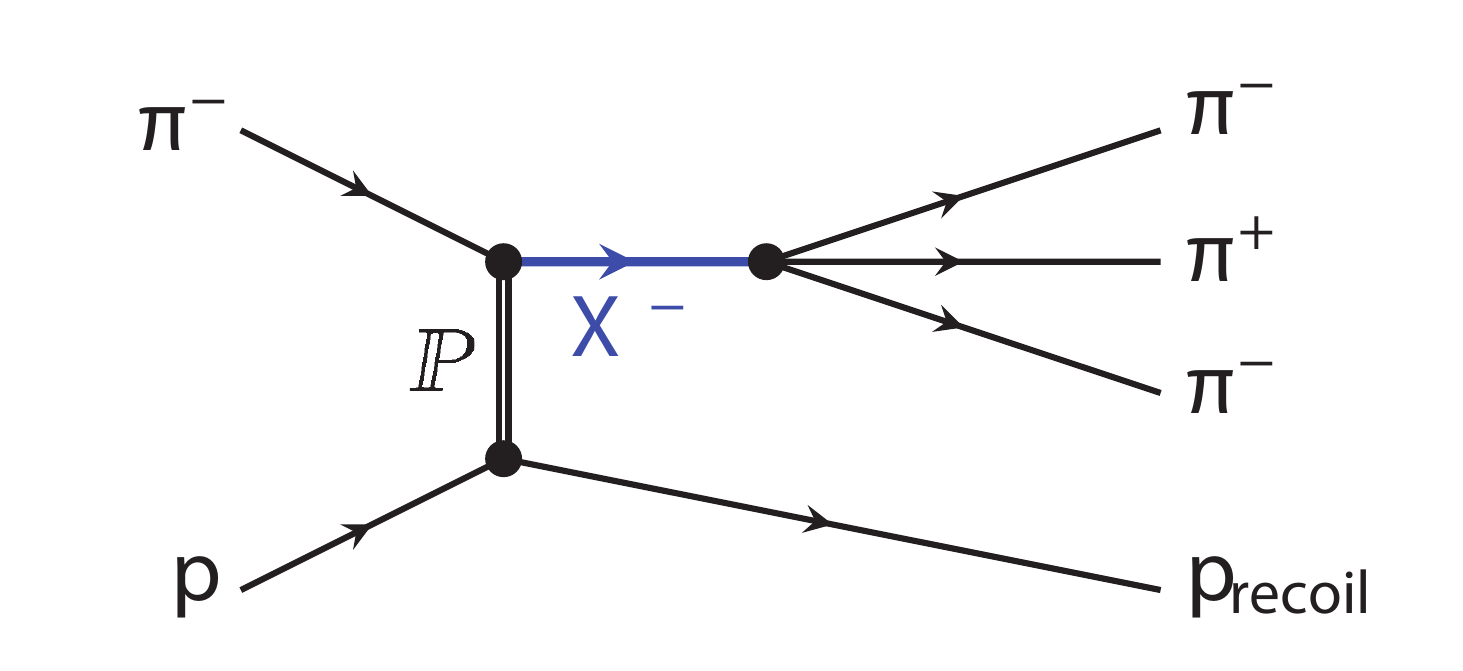}
\raisebox{-25pt}{\includegraphics[width=1.4in]{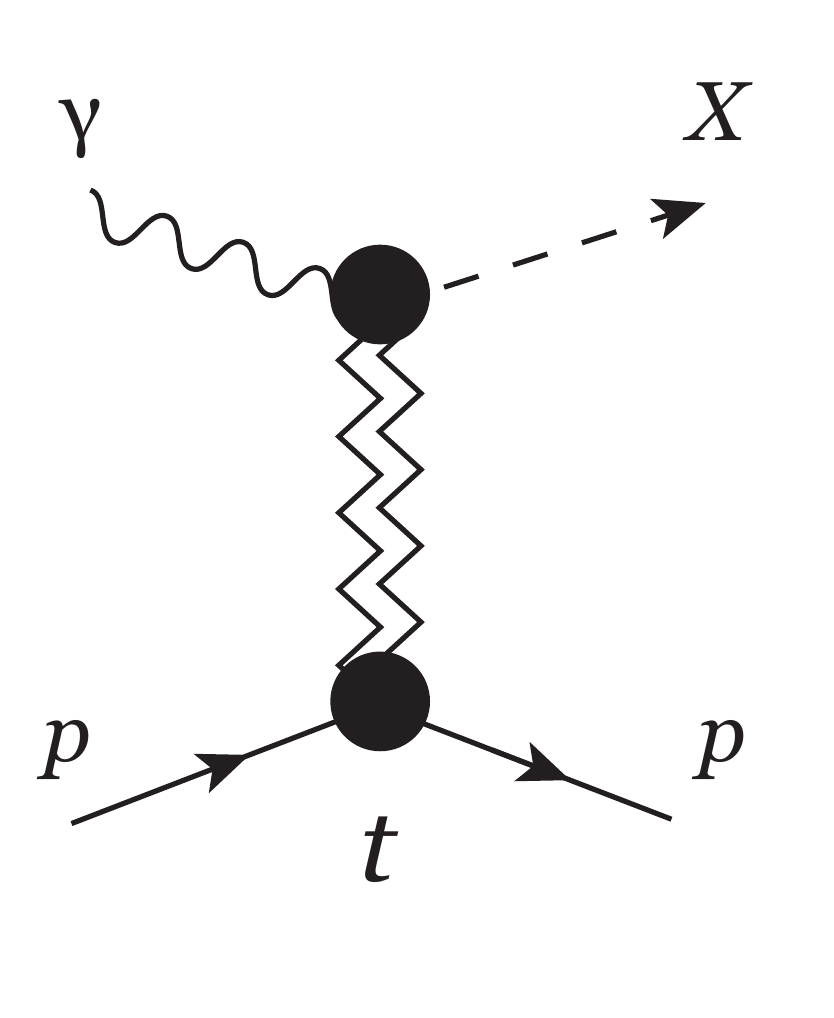}}
\end{center}

\caption{Examples of light meson production processes: (left) Production in hadron decay, in association with another particle, which can be selected to tune the properties of the produced light meson X; (middle) Diffractive hadroproduction from a high-energy pion beam, for the example of the production of a light meson $\mathrm{X}^-$ decaying to $\pi^-\pi^+\pi^-$, which is dominated by Pomeron exchange; (right)  Photoproduction off a proton target, where the quantum numbers of the light meson X depend on the quantum numbers of the quanta exchanged between the proton and the photon.}
\label{fig:production}
\end{figure}

However, the reality of observed states is more complicated than the simple constituent quark model, as more complex states with the same quantum numbers as normal mesons can quantum-mechanically mix with them to produce the states that are experimentally observed.  These more complex states include multiquark states and those with significant gluonic content.
There are then two important experimental challenges.  Most mesons have large intrinsic widths, and given the large number of meson states, multiple states will overlap with each other in the mass spectra for a given final state.  To identify individual states, amplitude analyses that can disentangle their separate contributions are required.  Then, to further identify the meson states and determine the contribution of different components to their wavefunctions, it is necessary to measure their decays to different final state particles, and to analyze these multiple decay channels in a consistent framework.  This program of analysis is making fresh progress in recent years thanks to large, high-quality data collected by modern, large-acceptance detectors, and improvements in theoretical models needed to understand these data.  In the following, I will discuss results from three of these new sets of data collected by different experiments using different reactions, which are illustrated in Fig.~\ref{fig:production}.

\section{Light Mesons at BES-III}

Studying mesons produced in the decay of other hadrons has several advantages.  It is often possible to select a clean sample of parent hadrons, yielding a sample of events with little background and a well defined initial state.  For two-body hadron decays, such as illustrated in Fig.~1(a), by selecting one of the particles produced in the decay to be of a particular type, one can select for the $J^{PC}$ and primary quark content of the other particle produced in the decay.  A classic example is the charmonium decay $J/\psi~(1^{--}) \to \gamma~(1^{--})  + \mathrm{X}~(0^{++}~\mathrm{and}~2^{++})$.  Instead of a photon, one could select an $\omega$ or $\phi$ meson to enhance the associated production of mesons with $ud$ and $s$ quark content, respectively.  This process has been utilized by many experiments, from CDF and D\O in high-energy $p\bar{p}$ collisions, and ATLAS, CMS, and LHCb in higher-energy $pp$ collisions, to BaBar, Belle, BES, CLEO, KLOE, and others in $e^+e^-$ annihilations.

BES-III is an experiment at the BEPC II $e^+e^-$ collider located at the Institute of High Energy Physics (IHEP) in Beijing, China.  The BES-III detector~\cite{besiii} is cylindrically symmetric and has reconstruction and particle identification capabilities to detect all particles produced in the $e^+e^-$ annihilations generated at $\sqrt{s} \sim 2-5$~GeV.  BES-III has collected the world's largest data samples in the charmonium region, consisting of $\sim 1.3\times10^9~J/\psi$, $\sim 5.0 \times 10^8 ~ \psi(2S)$, and $>6$~fb$^{-1}$ collected above $D\overline{D}$ thresholds.  These data have allowed for the study of light mesons with $M\lesssim2.5$~GeV in a wide variety of charmonium decays. 

%

\begin{figure}[!tb]
\begin{center}
\includegraphics[width=5.in]{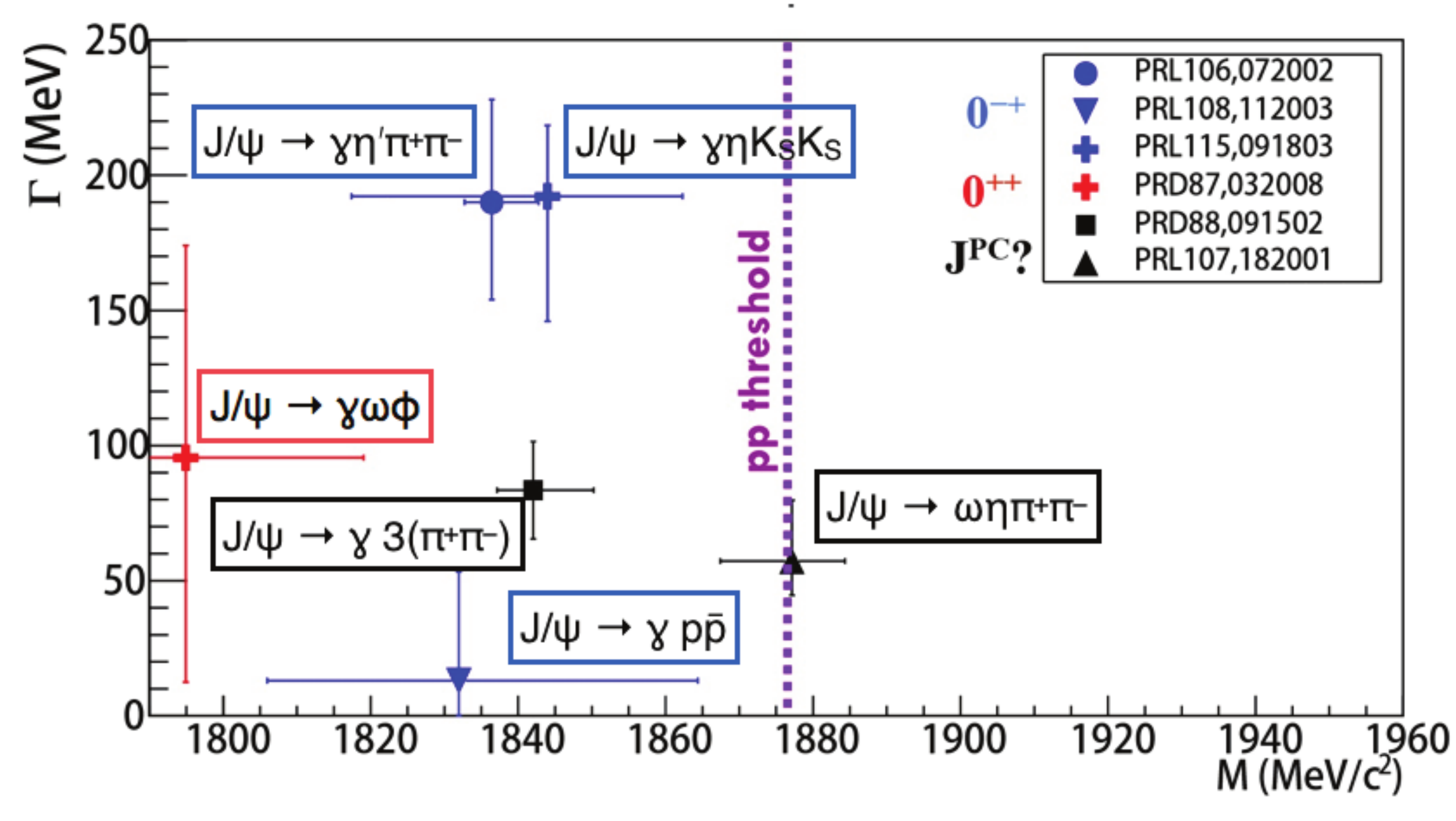}
\end{center}

\caption{Summary of the properties of the states identified by BES-III in the mass region $1790-1890$~MeV.  The axes give the Breit-Wigner masses and widths, and the $J^{PC}$ are given, when known.\cite{besiiippbar,besiiietappipi,besiii6pi,besiiietaksks,besiiiomegaeta,besiiiomegaetapipi}. Taken from Ref.~\cite{zhanghadron}.}
\label{fig:x1835}
\end{figure}

One of the biggest current mysteries raised by BES-III is the nature of the states that have been observed in the region near the $p\bar{p}$ threshold, $M\sim1800 - 1880$~MeV.  The first of the states was named X(1835), and was seen as a threshold enhancement in the $M(p\bar{p})$ distribution in the decay $J/\psi \to \gamma p\bar{p}$ by BES-II, and interpreted as a sub-threshold resonance~\cite{besiippbar}. A partial wave analysis (PWA) of this reaction with a larger data sample by BES-III identified its $J^{PC}$ as $0^{-+}$~\cite{besiiippbar}.  Subsequent studies of the decays of $J/\psi$ and $\psi(2S)$ to $(\gamma,\omega,\pi^0,\eta) p\bar{p}$ yielded little additional evidence for such a state~\cite{otherpp}.  However, in X,  a study of the decay $J/\psi \to \gamma \eta' \pi^+\pi^-$ found an enhancement in the $M(\eta' \pi^+\pi^-)$ spectrum with mass $\sim1835$~MeV, but a width of $\sim200$~MeV~\cite{besiietappipi,besiiietappipi}, much larger than the $<50$~MeV found for the X(1835) in the $p\bar{p}$ channel. 

Since then, several other states with various $J^{PC}$ have been found in this mass region.  Several of their mass spectra are shown in Fig.~2, and the current status of their resonance parameters and $J^{PC}$ is summarized in Fig.~\ref{fig:x1835}.  The main questions are now: how many states actually exist in this region and what is the relation of these different enhancements to each other?  The first step must be to firmly assign $J^{PC}$ values to all of these enhancements, either through PWA or other angular analysis.  Other complications to the interpretation of these states are the large backgrounds in some of the decay channels, which could affect the mass and width determinations, and the closeness of these states to the $p\bar{p}$ threshold.  These factors point towards the need for a coupled-channel analysis in order to accurately determine the lineshapes and resonance parameters of a possible state in these different decay channels.

\begin{figure}[!tb]
\begin{center}
\includegraphics[width=1.9in]{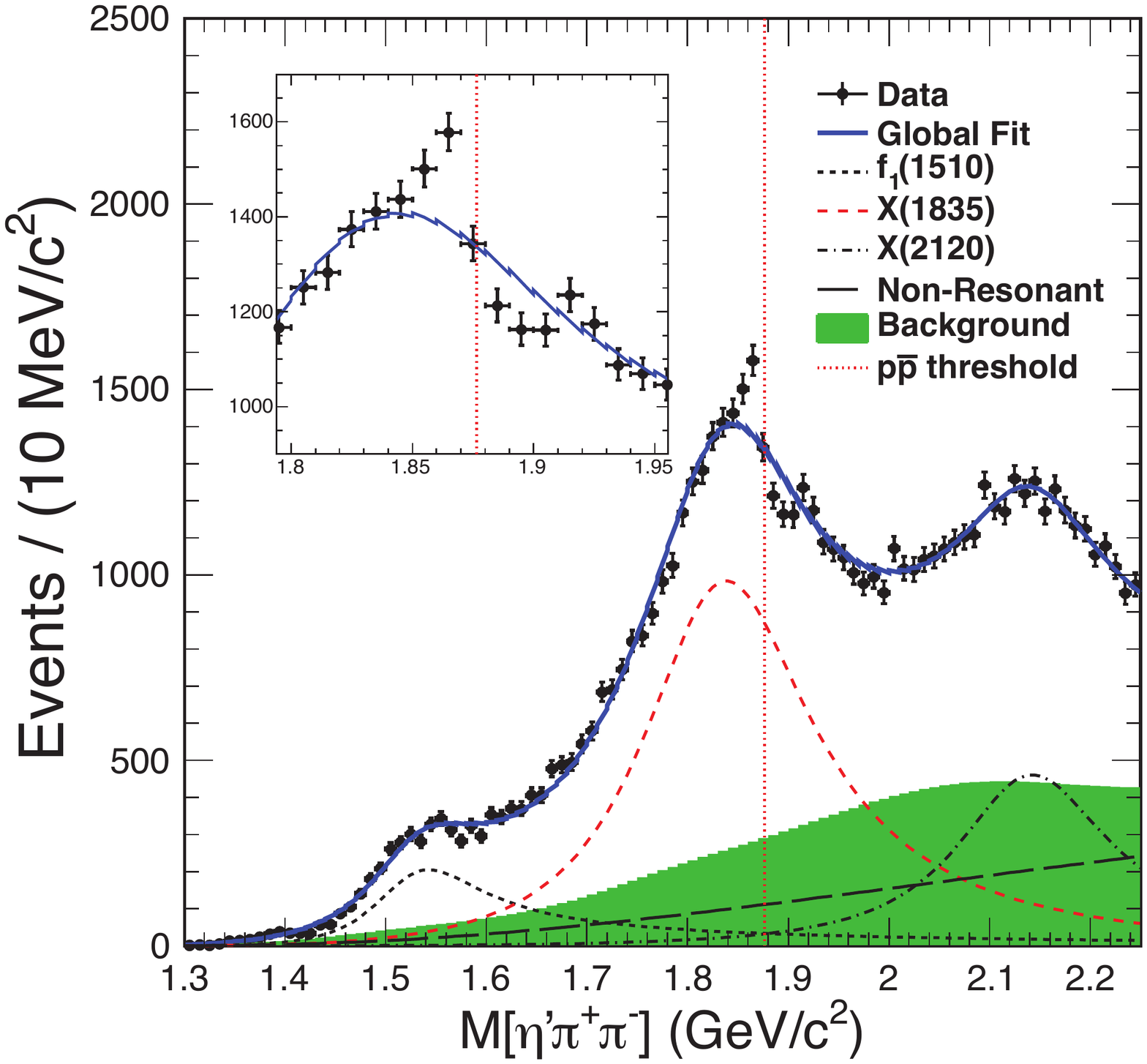}
\includegraphics[width=1.9in]{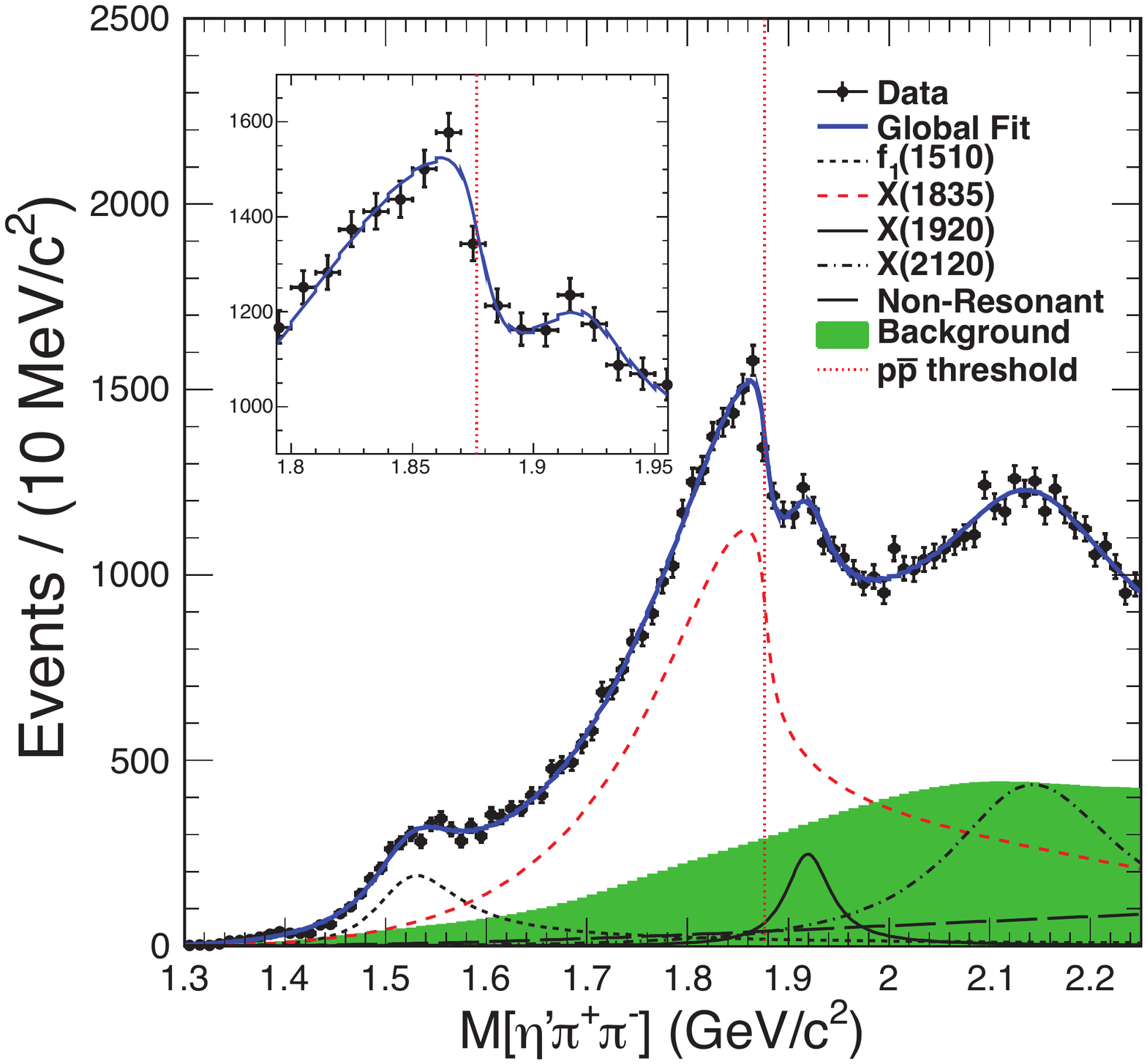}
\includegraphics[width=1.9in]{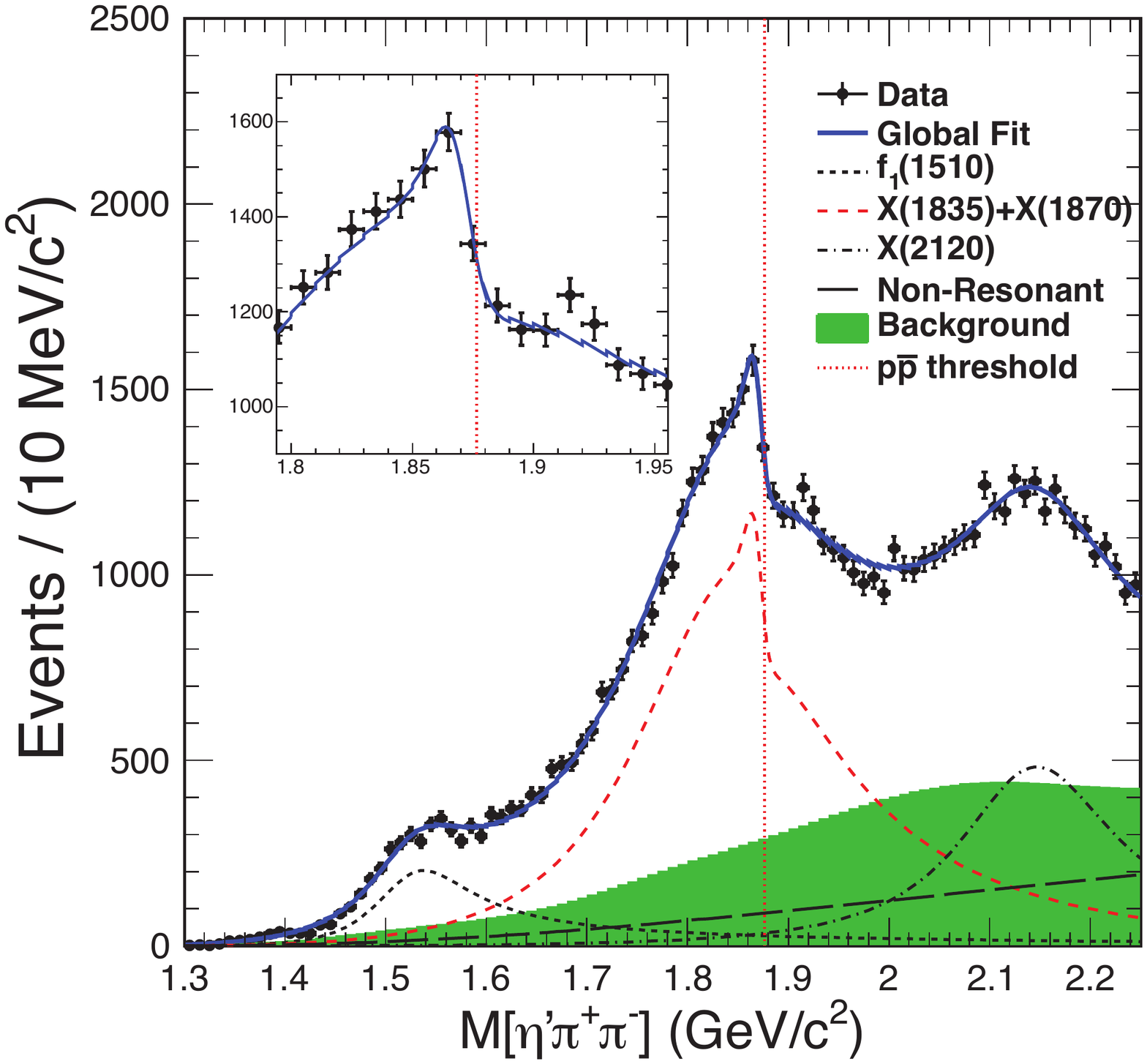}
\vspace*{-15pt}
\end{center}

\caption{Fits to the most recent $\eta'\pi^+\pi^-$ mass spectrum from BES-III, using different parameterizations for the peak at $M\sim1.8$~GeV.  (Left) A single Breit-Wigner shape, as used in the previous publication; (middle) a Flatte shape that is strongly coupled to the nearby $p\bar{p}$ threshold along with a narrow Breit-Wigner of mass $M\sim1.92$~GeV; (right) a coherent sum of two Breit-Wigners, a wide X(1835) and a narrow X(1870). From Ref.~\cite{besiiietappipimore}.}
\label{fig:prl_interfere}
\end{figure}

A first step towards a more sophisticated analysis was taken in the analysis of a larger data set for the reaction $J/\psi \to \gamma \eta' \pi^+\pi^-$~\cite{besiiietappipimore}.  Three fits to the $M(\eta' \pi^+\pi^-)$ spectrum are shown in Fig.~\ref{fig:prl_interfere}.  As opposed to the previous measurement, the peak at $M\sim1.8$~GeV is no longer described well by a simple Breit-Wigner function.  Results from two other models are shown: one with a Flatte shape that is strongly coupled to the nearby $p\bar{p}$ threshold along with a narrow Breit-Wigner of mass $M\sim1.92$~GeV, and a coherent sum of two Breit-Wigners, a wide X(1835) and a narrow X(1870).  Both models have a similar fit quality, and more information is needed to accurate describe the spectrum, again pointing in the direction of a coupled channel analysis.

\begin{figure}[!tb]
\begin{center}
\includegraphics[width=2.5in]{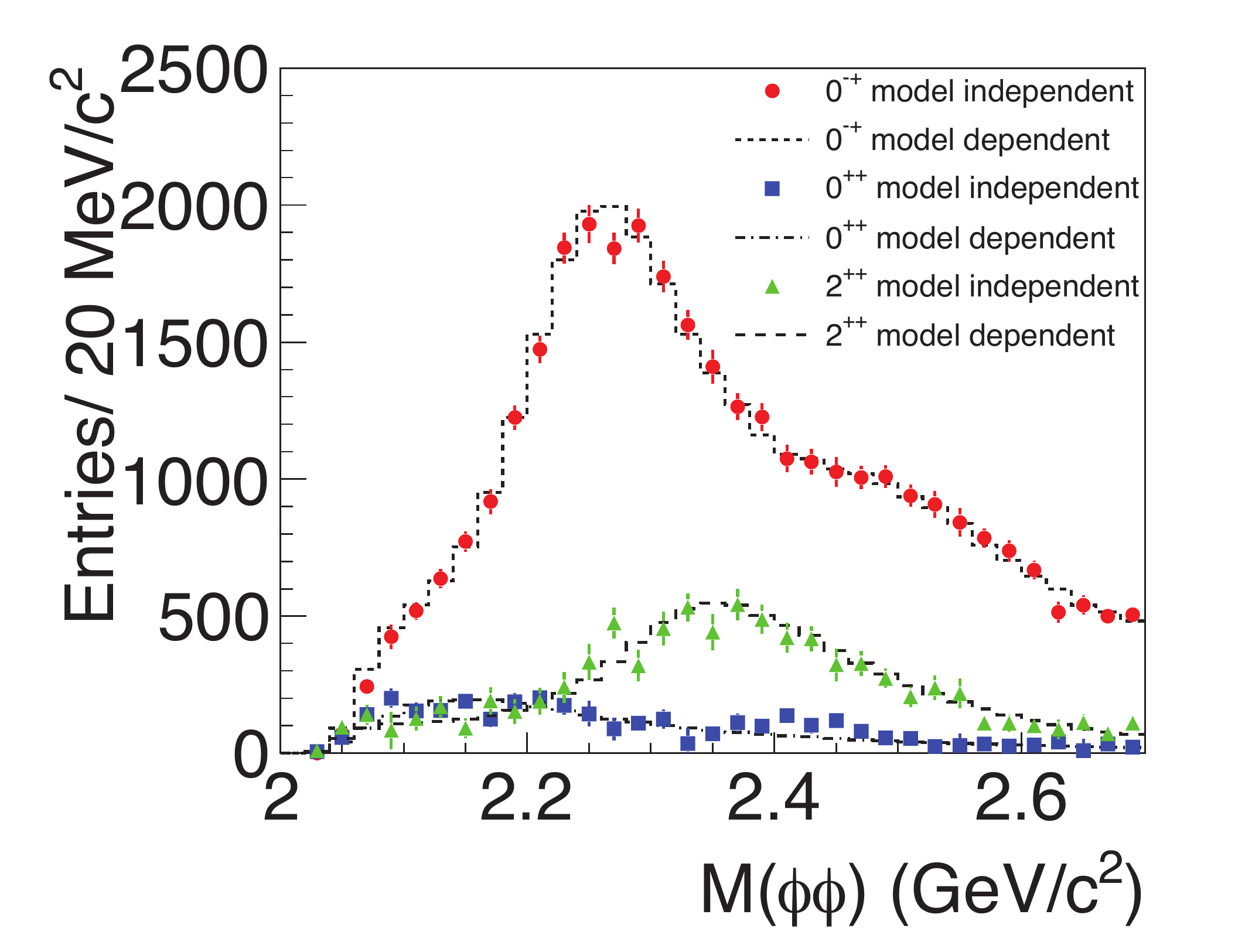}
\begin{scriptsize}
\raisebox{75pt}{
     \setlength{\tabcolsep}{4pt}
      \begin{tabular}{ccccc}
        \hline
        \hline
        Resonance     &M(MeV/$c^{2}$) &$\Gamma$(MeV/$c^{2}$) &B.F.($ \times 10^{-4}$)  &Sig.\\
        \hline
        $\eta(2225)$  &$2216_{-5}^{+4}$$_{-11}^{+21}$      &$185_{-14}^{+12}$$_{-17}^{+43}$    &$(2.40\pm0.10$$_{-0.18}^{+2.47})$         &28$\;\sigma$        \\
        $\eta(2100)$  &$2050_{-24}^{+30}$$_{-26}^{+75}$     &$250_{-30}^{+36}$$_{-164}^{+181}$ &$(3.30\pm0.09$$_{-3.04}^{+0.18})$         &22$\;\sigma$      \\
        $X(2500)$       &$2470_{-19}^{+15}$$_{-23}^{+101}$      &$230_{-35}^{+64}$$_{-33}^{+56}$  &$(0.17\pm0.02$$_{-0.08}^{+0.02})$         &8.8$\;\sigma$\\
        $f_{0}(2100)$ &2101                     &224                                           &$(0.43\pm0.04$$_{-0.03}^{+0.24})$         &24$\;\sigma$\\
        $f_{2}(2010)$ &2011                     &202                                           &$(0.35\pm0.05$$_{-0.15}^{+0.28})$         &9.5$\;\sigma$ \\
        $f_{2}(2300)$ &2297                     &149                                           &$(0.44\pm0.07$$_{-0.15}^{+0.09})$         &6.4$\;\sigma$ \\
        $f_{2}(2340)$ &2339                     &319                                           &$(1.91\pm0.14$$_{-0.73}^{+0.72})$         &11$\;\sigma$\\
        $0^{-+}$ PHSP &                         &                                              &$(2.74\pm0.15$$_{-1.48}^{+0.16})$         &6.8$\;\sigma$\\
        \hline
        \hline
      \end{tabular}
      }
\end{scriptsize}

\end{center}

\caption{Summary of results from the PWA of the reaction $J/\psi \to \gamma \phi\phi$ from BES-III.  The left panel shows the mass spectrum and contributions from the different partial waves.  The right gives the results for the different resonant contributions in tabular form.  From Ref.~\cite{besiiiphiphi}.}
\label{fig:phiphi}
\end{figure}

Several PWAs of radiative $J/\psi$ decays have also been performed.  The most recently published result was of the PWA of the $J/\psi \to \gamma \phi\phi$ decay~\cite{besiiiphiphi}.  This reaction allowed for the study of scalar, pseudoscalar, and tensor mesons in the little-studied region of $M>2.0$ GeV.  The results are summarized in Fig.~\ref{fig:phiphi}.  This analysis confirmed the contribution of several $f_2$ states and the $\eta(2225)$, and identified two new states, the $\eta(2100)$ and X(2500).  Several other PWA's are ongoing.  The results of the PWA of the radiative production two pseudoscalar mesons is particularly interesting for the information it gives on the spectrum of scalar mesons and the potential contribution of glueballs.  The results for $J/\psi \to \gamma \pi^0\pi^0$~\cite{besiiipi0pi0} and $J/\psi \to \gamma \eta\eta$~\cite{besiiietaeta} have been already published, and the results of the analysis of the $\pi^+\pi^-$, $K^+K^-$, and $K_SK_S$ final states are eagerly awaited.

\section{Light Mesons at COMPASS}

The study of mesons produced using beams of hadrons (primarily charged $\pi$ and $K$ mesons) has several advantages, including the ability to tune the beam particle type in order to preferentially create different types of hadrons, and the ability to collect large sets of data, since hadron beam experiments are almost always fixed-target experiments. The primary downside of such hadroproduction experiments, compared to the meson production in decays discussed in the previous section, is that their analysis is more complicated.  Instead of starting from an initial state with well-defined quantum numbers, the final states observed in hadroproduction experiments can be produced through several different processes.  Untangling their contributions for the study of mesons requires the application of appropriate models.  Due to the ease of production of pion beams, meson pionproduction has been studied in many experiments, including E852 at Brookhaven, VES, and COMPASS.

\begin{figure}[!tb]
\begin{center}
\includegraphics[width=5.in]{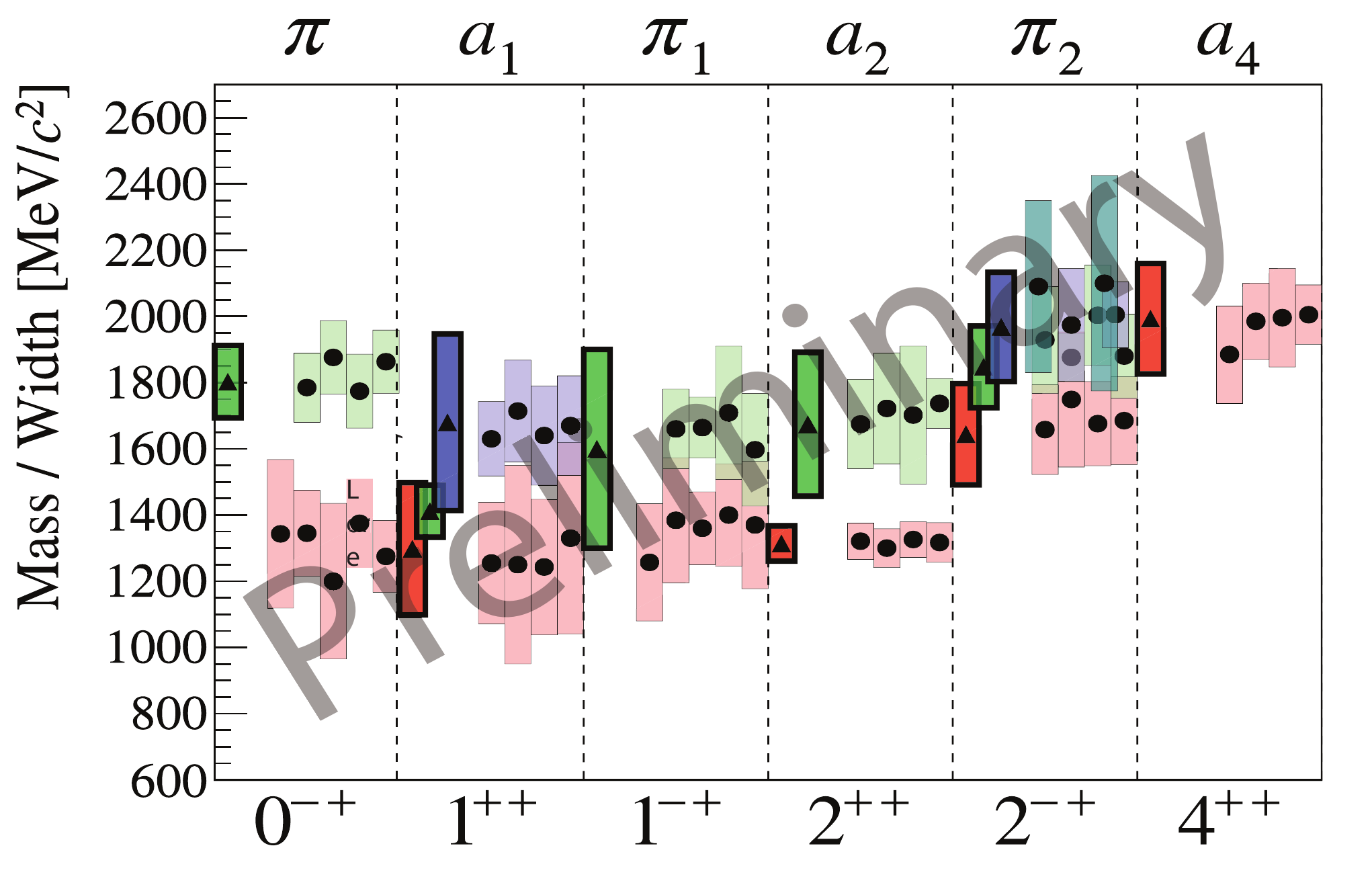}
\end{center}

\caption{Summary of preliminary results of the analysis of $\sim50\times10^6$ exclusive $\pi^- + p \to \pi^-\pi^+\pi^- + p_\mathrm{recoil}$ events.
Center of the boxes represents the mass, height of the boxes the width of the states. The different colors show ground and excited states. The circles represent the latest measurements according to PDG 2014, the triangles the results of this analysis.}
\label{fig:compasssummary}
\end{figure}

The COMPASS experiment~\cite{compass} is a fixed target experiment located at CERN which can support a variety of muon and hadron beams.  Data were taken in several periods with a 190~GeV/$c$ $\pi^-$ beam incident on a liquid hydrogen target in order to study the light meson spectrum with $M\lesssim2$~GeV using diffractive pion-proton scattering (see Fig.~1(middle)).   Several different final states including $\pi$, $\eta$, and $\eta'$ mesons have been investigated.  Notably, COMPASS has a collected a large sample of $\sim50\times10^6$ exclusive $\pi^- + p \to \pi^-\pi^+\pi^- + p_\mathrm{recoil}$ events.  A $t-$resolved analysis in bins of momentum transfer has been performed, using an impressive 88 partial waves, the largest set to date.  The results of this partial wave analysis have been already published~\cite{compasspwa}.

\begin{figure}[!tb]
\begin{center}
\includegraphics[width=4.5in]{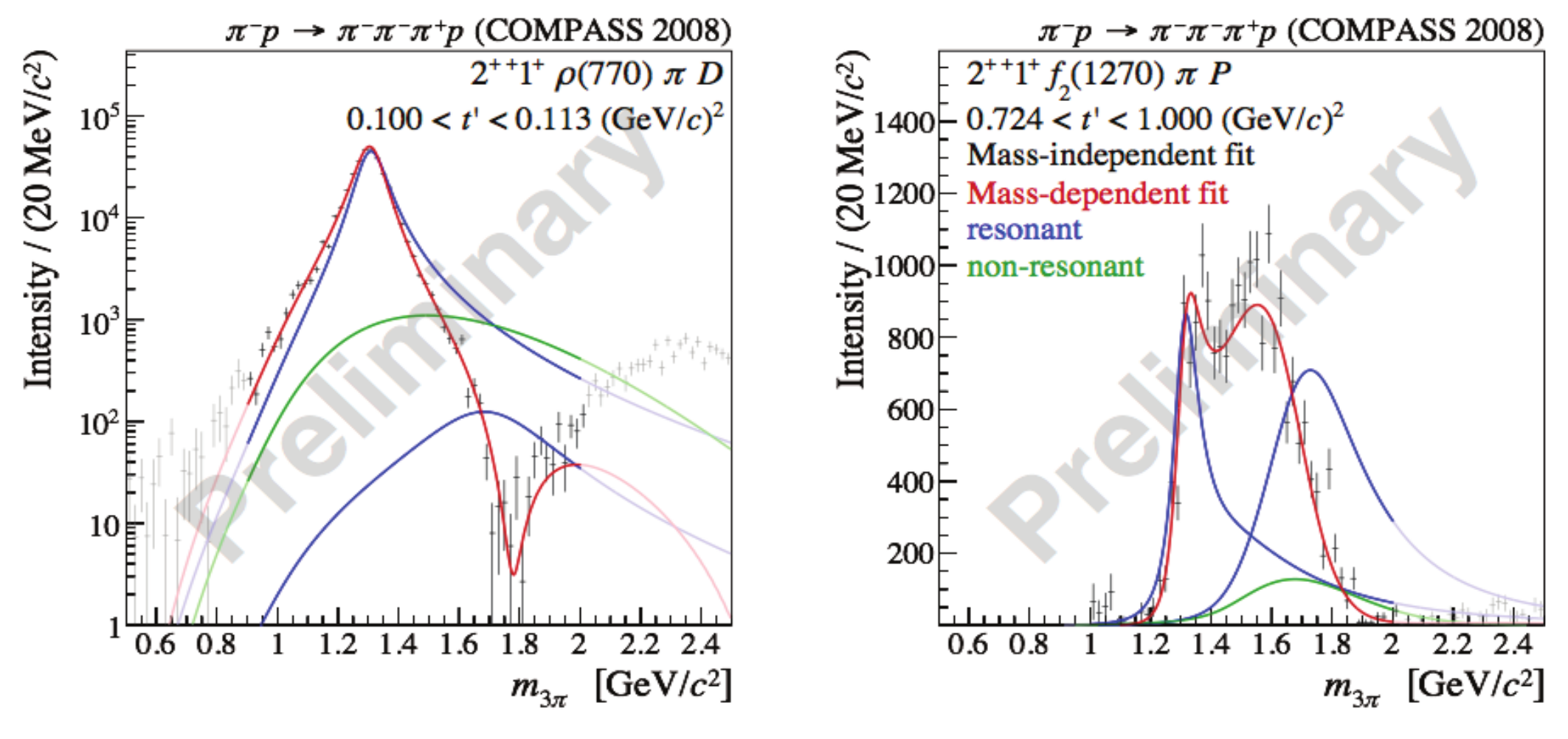}
\end{center}

\caption{Preliminary fits to COMPASS data showing contributions from the $2^{++}$ $a_2$ mesons.}
\label{fig:compassa2}
\end{figure}

\begin{figure}[!tb]
\begin{center}
\includegraphics[width=4.5in]{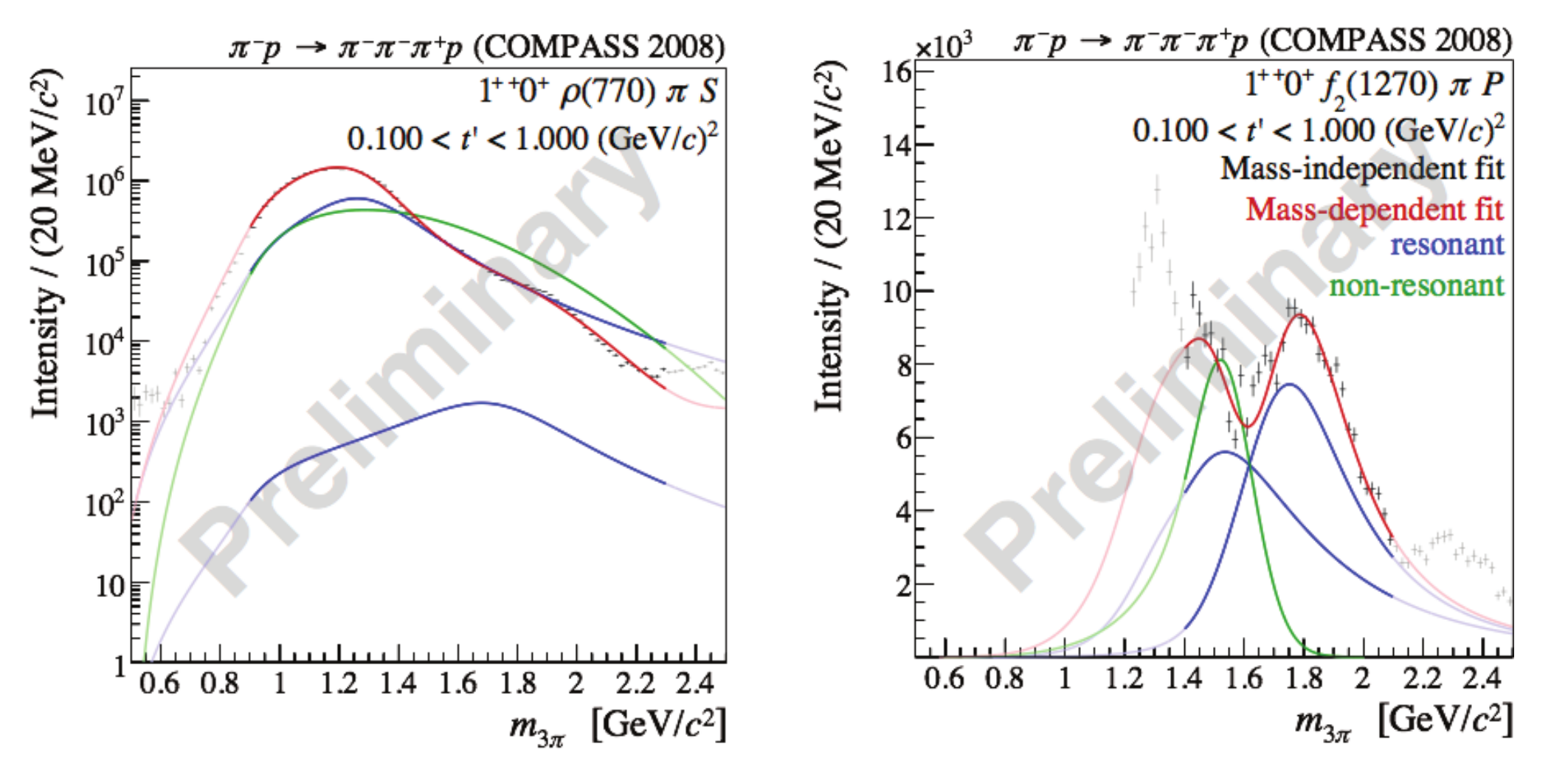}
\end{center}

\caption{Preliminary fits to COMPASS data showing contributions from $1^{++}$ $a_1$ mesons.}
\label{fig:compassa1}
\end{figure}

\begin{figure}[!tb]
\begin{center}
\includegraphics[width=5.5in]{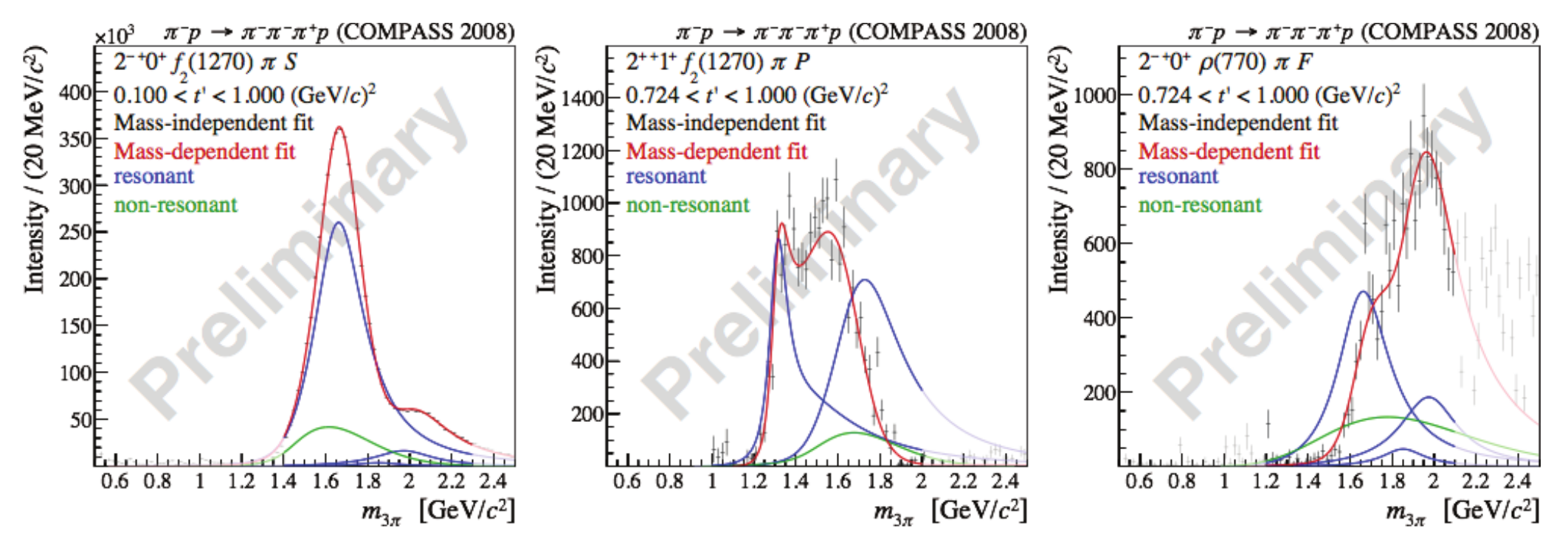}
\end{center}

\caption{Preliminary fits to COMPASS data showing contributions from $2^{-+}$ $\pi_2$ mesons.}
\label{fig:compasspi2}
\end{figure}

The next step in the analysis of this reaction is to perform a resonance model fit to this data in order to determine the contributions of different intermediate states in this reaction and their properties.  The current status of this model fit is discussed in detail elsewhere in these proceedings~\cite{compassnew}, but briefly it models the $M(\pi^-\pi^+\pi^-)$ dependence using  resonant and non-resonant contributions from 11 ground and excited states in a simultaneous fit to 14 partial waves.  This is the largest model used in such an analysis so far, and extensive systematic studies have been done with this fit.

The preliminary results from this fit are summarized in Fig~\ref{fig:compasssummary}, and results from individual partial waves are shown in Figs.~\ref{fig:compassa2}, \ref{fig:compassa1}, and \ref{fig:compasspi2}.  In the $2^{++}$ channels, a robust signal for the well-known $a_2(1320)$ is found, as illustrated in Fig.~\ref{fig:compassa2}, while evidence for the lesser-known $a_2(1700)$ is seen as destructive interference at low $t'$, particularly in the $f_2(1270)\,\pi$ P-wave channel.  In the $1^{++}$ channels, evidence for a substantial non-resonant contribution is seen, for example the primary contributions to the  $\rho(770)\,\pi$ S-wave are found to be the well-known $a_1(1260)$ and a non-resonant contribution of similar size, as illustrated in Figs.~\ref{fig:compassa1}.  A potential signal for the $a_1(1640)$ is also seen, with the strongest evidence in the $f_2(1270)\,\pi$ P-wave channel.  Figs.~\ref{fig:compasspi2} illustrates the evidence for the well-known $\pi_2(1670)$, the lesser-known $\pi_2(1880)$, and a new $\pi_2(2005)$.  To summarize, COMPASS has made robust measurements of the properties of the ground states for several $J^{PC}$ mesons, and has provided valuable measurements for several excited states, for which few measurements currently exist.

The further analysis of this COMPASS data is expected to continue to provide a wealth of knowledge on light quark states.  Ongoing projects related to the $\pi^-\pi^+\pi^-$ final state include the extraction of resonance contributions to the $\pi^+\pi^-$ subsystem, and analysis of models and partial waves using ``semi-automatic'' algorithmical methods~\cite{compassfuture}.  Other non-strange final states are also being studied, and collaborations with other groups are leading to analysis that move beyond the standard isobar model.  One such example is the collaboration with JPAC, which yielded an analysis including analyticity and unitarity constraints~\cite{compassjpac}.  There is also the possibility of studying strange meson final states using data collected with a charged kaon beam.

\section{Light Mesons at GlueX}

The study of light mesons in photoproduction has gained new interest with the availability of multi-GeV photon beams at Jefferson Lab.  In meson photoproduction, as illustrated in the right panel of Fig.~1, the photon couples through vector meson dominance (VMD) with quanta exchanged from the target (usually a proton) which are generally modeled as the exchange of a virtual meson.  The wide variety of possible couplings allows for the production of a wide variety of mesons of different quark content and quantum number.  Conversely, this flexibility means that more complicated theoretical models are generally needed to describe these interactions than for hadroproduction.

The GlueX experiment~\cite{gluex} is the flagship experiment for the newly constructed Hall D  in Jefferson Lab located in Newport News, Virginia.  GlueX takes the highest-energy electrons extracted from the upgrade 12 GeV CEBAF electron beam accelerator and scatters them off of a thin diamond radiator, to create a broadband photon beam peaked at 9~GeV, with a high degree of linear polarization.  The photon beam is incident on a liquid hydrogen target, and is surrounded by a spectrometer with good capabilities to detect both charged and neutral particles.  The experiment had a commissioning run in 2016, and has started a multi-year program of data taking in early 2017.  
The primary goal of GlueX is the identification and study of the spectrum of hybrid mesons, which are mesons where the confining gluonic field contributions directly to the properties of the meson.  The study of the spectrum of light mesons is a prerequisite to these exotic searches, and the data taken from GlueX can be used for many studies of hadronic physics.
The GlueX detector and physics program are discussed in more detail elsewhere in these proceedings~\cite{gluexhadron}, but I will discuss a few highlights below.

\begin{figure}[!tb]
\begin{center}
\raisebox{5pt}{\includegraphics[width=2.45in]{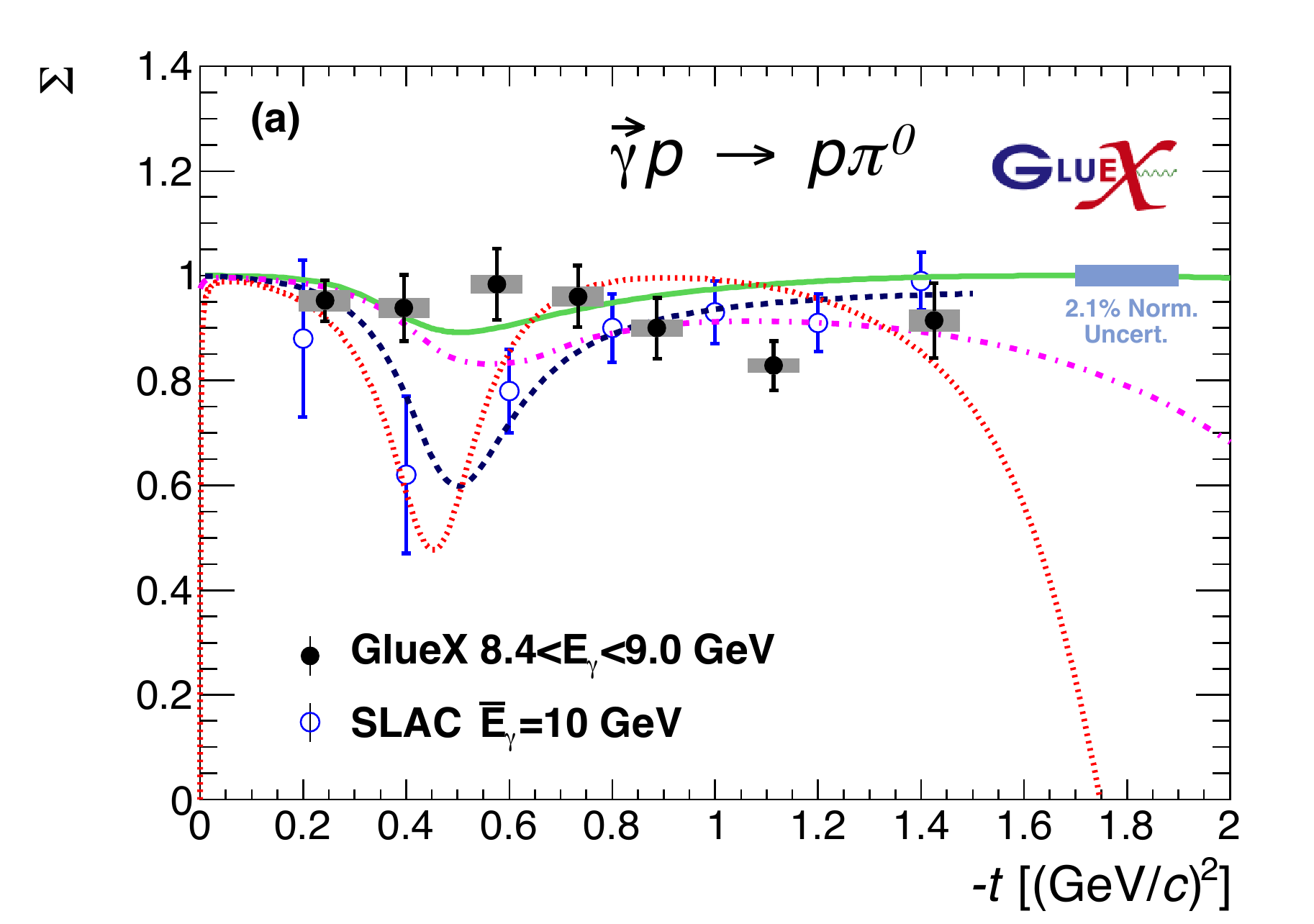}}
\includegraphics[width=2.5in]{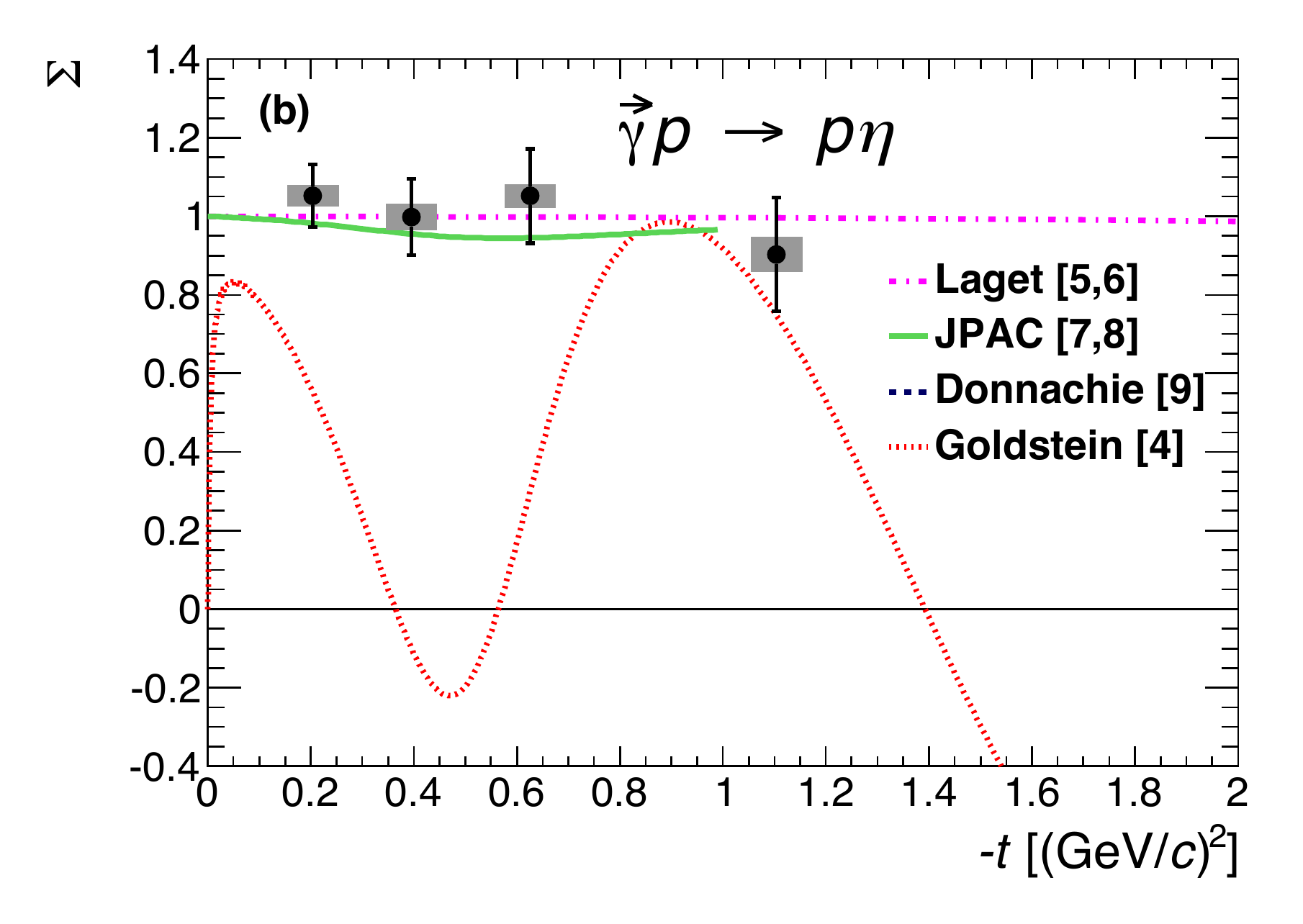}
\end{center}

\caption{Beam asymmetry $\Sigma$ from 2016 GlueX data as a function of Mandelstam$-t$ for the production of $\pi^0$ (middle) and $\eta$ (right), along with several model predictions. From Ref.~\cite{gluexpi0}.}
\label{fig:gluexsigma}
\end{figure}

Photoproduction near 9 GeV has been little studied in many years, with no new experimental data since the SLAC experiments of the 1970s and early 1980s.  The first step towards the GlueX program of studying the light meson spectrum and searching for hybrid mesons is necessarily to study the processes through which they are produced.  These studies will yield important inputs into the amplitude analyses necessary for meson spectroscopy, and to do so requires working closely with theoretical colleagues, notably the JPAC Collaboration.  Such collaborations are already yielding new models for interpreting the data coming out of GlueX, which bodes well for the future program of photoproduction studies.

To begin the study of photoproduction processes, a program of studying the beam asymmetries ($\Sigma$) for the photoproduction of single psuedoscalar mesons ($\pi^0,\pi^-,\eta,\eta',...$) and the spin-density matrix elements for the photoproduction of single vector mesons ($\rho,\omega,\phi,...$) has begun.  Preliminary results for many of these channels have been recently presented at conferences, and the first measurements of the beam asymmetry in $\pi^0$ and $\eta$ photoproduction using the 2016 data set has resulted in the first GlueX publication~\cite{gluexpi0}.  These results are illustrated in Fig.~\ref{fig:gluexsigma}.  The value of $\Sigma$ of near unity indicates the dominance of vector meson exchange in this process.  The $\eta$ measurement is the first in this energy range, and these results illustrate the first steps towards a detailed understanding of the production mechanisms of mesons in this energy range.

\begin{figure}[!tb]
\begin{center}
\includegraphics[width=6.in]{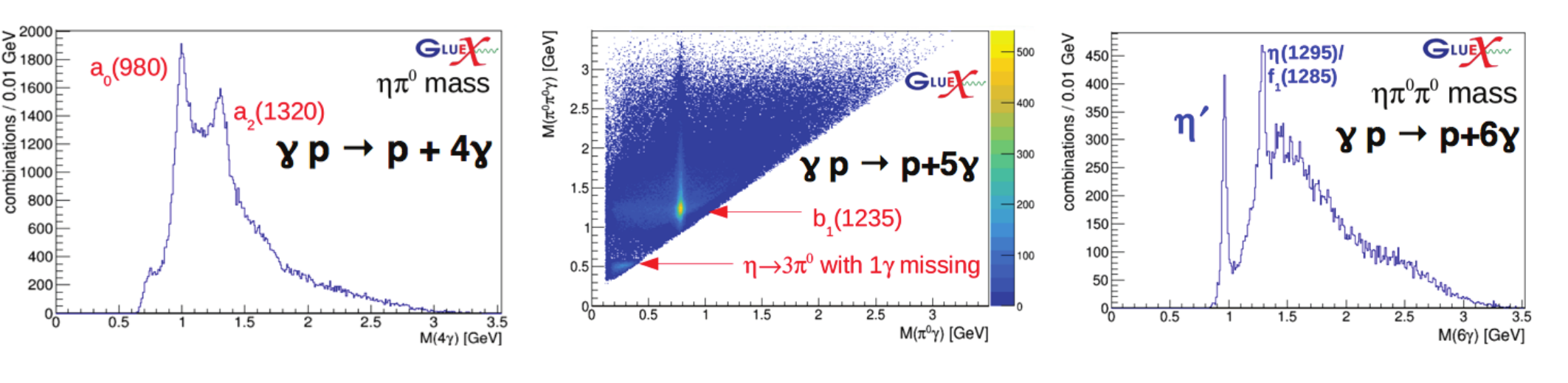}
\end{center}

\caption{Mass spectra from exclusive $\gamma p \to p + (4,5,6)\gamma$ events in early GlueX data. Contributions likely due to well-known states are labeled.}
\label{fig:gluexsigma}
\end{figure}

\begin{figure}[!tb]
\begin{center}
\includegraphics[width=6.in]{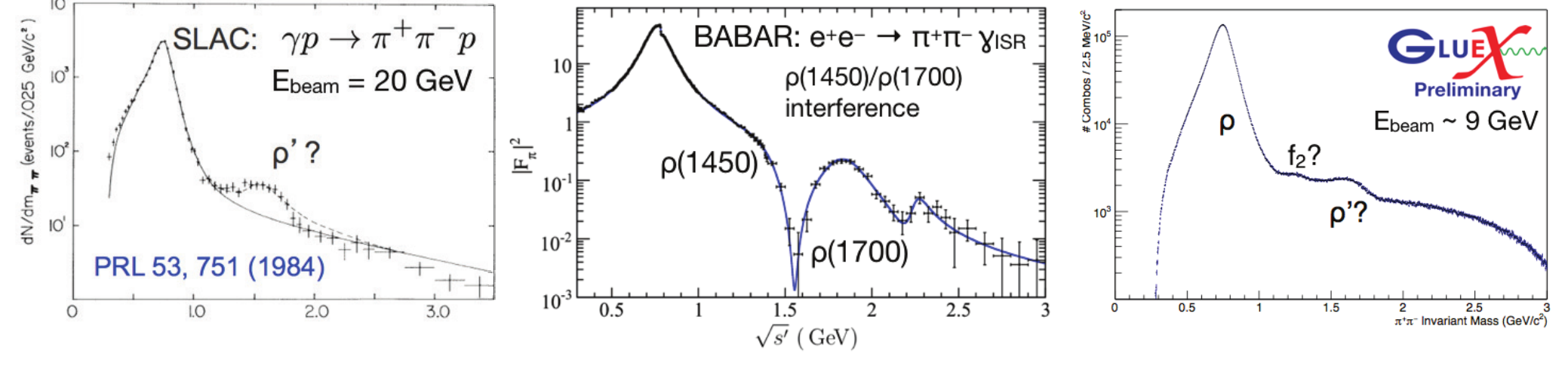}
\end{center}

\caption{(Left) Invariant mass spectrum of $\pi^+\pi^-$ from exclusive $\gamma p \to \pi^+\pi^-p$ events measured at SLAC, illustrating claimed $\rho(1600)$ contribution~\cite{slacrho}. (Middle) Timelike pion form factors determined by BaBar in $e^+e^- \to \gamma_{\mathrm{ISR}} \pi^+\pi^-$~\cite{babarpipi}.  The solid line illustrates the fit to the data with various $\rho$ resonances. (Right) Preliminary $\pi^+\pi^-$ invariant mass spectrum from early GlueX data illustrating various potential resonance contributions.}
\label{fig:gluexsigma}
\end{figure}

Final states containing neutral particles are almost unexplored in this energy range.  As an illustration of the prospects for the analysis of such reactions with GlueX, example mass spectra for the reactions $\gamma p \to p + (4,5,6)\gamma$ from initial GlueX data are shown in Fig.~X.  Many well-known mesons are clearly seen in this data. With at least an order of magnitude more data expected, the prospects for spectroscopy from amplitude analyses of these reactions look promising.. 

An example of the prospects for studying charged particle final states can be seen in $\gamma p \to p + \pi^+\pi^-$.  A basic question for understanding the light meson spectrum is, what is the spectrum of excited $\rho$ mesons?  Several candidates for excited $\rho$'s have been seen over the years, including the observation of a state called the $\rho(1600)$ in photoproduction in SLAC in 1984, shown in Fig.~X(left)~\cite{slacrho}.  However, the analysis of other reactions and other final states (notably the $4\pi$ final state) have led to the conclusion recorded in the PDG that there are most likely two different states in this mass region, the $\rho(1450)$ and $\rho(1700)$.  Many of these data have limited statistical precision, however, one high-statistics analysis is illustrated in Fig.~X(middle).  The BaBar data for $e^+e^- \to \gamma_\mathrm{ISR} \pi^+\pi^-$ around 1.5~GeV is found to be best described by the interference of the $\rho(1450)$ and $\rho(1700)$~\cite{babarpipi}.  

These studies then beg the question:  what is going on in photoproduction?  Are different states being produced, or are the $\rho(1450)$ and $\rho(1700)$ simply manifesting differently in the photoproduced $\pi^+\pi^-$ spectrum due to the different production processes that contribute, compared to other reactions.  GlueX can already take another look at this reaction, with two orders of magnitude more data than available at SLAC.  In a first look at the $\pi^+\pi^-$ mass spectrum, shown in Fig.~X(right), several enhancements are indeed seen with $M(\pi^+\pi^-) > 1$~GeV.  Moment and amplitude analyses are underway to determine their nature.

\section{Summary}

Although the study of light mesons has a venerable history, their study has been reinvigorated in recent years.  We are entering the era of large, high-quality data sets, and confronting the challenge of understanding these precise data is leading the way towards a better understanding of the light meson spectrum.  We can expect more data and refined analyses from BES-III and COMPASS in the coming years.  Besides the start of GlueX data taking, we can also look forward to data from the CLAS12 experiment at Jefferson Lab, which will use both electron and photon beams, and the PANDA $p\bar{p}$ annihilation experiment at GSI/FAIR, currently slated to start running no earlier than 2022.  With renewed efforts from both experiment and theory working in close collaboration, the future of this field looks bright indeed!


\begin{thebibliography}{99}
  \bibitem{review} For a general review of the subject, see, e.g., S. Godfrey and J. Napolitano, Rev.~Mod.~Phys. \textbf{71}, 1411 (1999).
  \bibitem{pdg} C. Patrignani et al. [Particle Data Group], Chin. Phys. C, \textbf{40}, 100001 (2016) and 2017 update.
  \bibitem{lattice} For example, J. J. Dudek, R. G. Edwards, P. Guo, and C. E. Thomas [Hadron Spectrum Collaboration], Phys. Rev. D \textbf{88}, 094505 (2013).
  \bibitem{besiii} M. Ablikim et al., [BES-III Collaboration], Nucl. Instr. Meth. Phys. A \textbf{614},  345 (2010).
  \bibitem{besiippbar} J.~Z.~Bai et al. [BES-II Collaboration], Phys. Rev. Lett. \textbf{91}, 022001 (2003).
  \bibitem{besiiippbar} M. Ablikim et al. [BES-III Collaboration], Phys. Rev. Lett. \textbf{108}, 112003 (2012).
  \bibitem{besiietappipi}  M. Ablikim et al. [BES-II Collaboration],  Phys. Rev. Lett. \textbf{95}, 262001 (2005). 
  \bibitem{besiiietappipi} M. Ablikim et al. [BES-III Collaboration], Phys. Rev. Lett. \textbf{106}, 072002 (2011).
  \bibitem{besiii6pi} M. Ablikim et al. [BES-III Collaboration],  Phys. Rev. D \textbf{88}, 091502(R) (2013).
  \bibitem{besiiietaksks} M. Ablikim et al. [BES-III Collaboration], Phys. Rev. Lett. \textbf{115}, 091803 (2015).
  \bibitem{besiiiomegaeta} M. Ablikim et al. [BES-III Collaboration], Phys. Rev. D \textbf{87}, 032008 (2013).
  \bibitem{besiiiomegaetapipi} M. Ablikim et al. [BES-III Collaboration], Phys. Rev. Lett. \textbf{107}, 182001 (2011).
  \bibitem{otherpp} M. Ablikim et al. [BES-II Collaboration], Eur. Phys. J. C \textbf{53}, 15 (2008);  M. Ablikim et al. [BES-II Collaboration], Phys. Rev. D \textbf{80}, 052004 (2009); J. P. Alexander, et al. [CLEO Collaboration], Phys. Rev. D \textbf{82}, 092002 (2010); M. Ablikim et al. [BES-III Collaboration], Phys. Rev. Lett. 110, 022001 (2013).
  \bibitem{besiiietappipimore} M. Ablikim et al., [BES-III Collaboration], Phys. Rev. Lett. \textbf{117}, 042002 (2016).
  \bibitem{zhanghadron} J. Zhang, elsewhere in these proceedings.
  \bibitem{besiiiphiphi} M. Ablikim et al., [BES-III Collaboration], Phys. Rev. D \textbf{93}, 112011 (2016).
  \bibitem{besiiipi0pi0} M. Ablikim et al., [BES-III Collaboration], Phys. Rev. D \textbf{92}, 052003 (2015).
  \bibitem{besiiietaeta} M. Ablikim et al. [BESIII Collaboration], Phys. Rev. D \textbf{87}, 092009 (2013).
  
  \bibitem{compass} P. Abbon, et al. [COMPASS Collaboration], Nucl. Instrum. and Meth. A \textbf{577}, 455 (2007).
  \bibitem{compasspwa} C. Adolph et al. [COMPASS Collaboration], 	Phys. Rev. D \textbf{95}, 032004 (2017).
  \bibitem{compassnew} S, Wallner, elsewhere in these proceedings.
  \bibitem{compassfuture} F. Krinner and B. Grube, elsewhere in these proceedings.
  \bibitem{compassjpac} A. Jackura et al. [COMPASS and JPAC Collaborations], \texttt{arXiv:1707.02848 [hep-ph]}.
  
  \bibitem{gluex} H. Al Ghoul et al. [GlueX Collaboration], AIP Conf. Proc. \textbf{1735} 020001 (2016).
  \bibitem{gluexhadron} S. Dobbs, elsewhere in these proceedings.
  \bibitem{gluexpi0} H. Al Ghoul et al. [GlueX Collaboration], Phys. Rev. C \textbf{95}, 042201 (2017).
  \bibitem{slacrho} K. Abe et al. [SLAC Hybrid Facility Photon Collaboration], Phys. Rev. Lett. \textbf{53}, 751 (1984).
  \bibitem{babarpipi}  J. P. Lees et al. [BaBar Collaboration], Phys. Rev. D \textbf{86} 032013 (2012).

\end{thebibliography}
\end{document}